\documentclass[twoside,twocolumn,9pt]{article}
\usepackage{extsizes}
\usepackage[super,sort&compress,comma]{natbib} 
\usepackage[version=3]{mhchem}
\usepackage[left=1.5cm, right=1.5cm, top=1.785cm, bottom=2.0cm]{geometry}
\usepackage{balance}
\usepackage{mathptmx}
\usepackage{sectsty}
\usepackage{graphicx} 
\usepackage{lastpage}
\usepackage[format=plain,justification=justified,singlelinecheck=false,font={stretch=1.125,small,sf},labelfont=bf,labelsep=space]{caption}
\usepackage{float}
\usepackage{fancyhdr}
\usepackage{fnpos}
\usepackage[english]{babel}
\addto{\captionsenglish}{%
  
}
\usepackage{array}
\usepackage{droidsans}
\usepackage{charter}
\usepackage[T1]{fontenc}
\usepackage[usenames,dvipsnames]{xcolor}
\usepackage{setspace}
\usepackage[compact]{titlesec}
\usepackage{hyperref}

\usepackage{epstopdf}

\definecolor{cream}{RGB}{222,217,201}

\begin{document}

\pagestyle{fancy}
\thispagestyle{plain}
\fancypagestyle{plain}{
\renewcommand{\headrulewidth}{0pt}
}

\makeFNbottom
\makeatletter
\renewcommand\LARGE{\@setfontsize\LARGE{15pt}{17}}
\renewcommand\Large{\@setfontsize\Large{12pt}{14}}
\renewcommand\large{\@setfontsize\large{10pt}{12}}
\renewcommand\footnotesize{\@setfontsize\footnotesize{7pt}{10}}
\makeatother

\renewcommand{\thefootnote}{\fnsymbol{footnote}}
\renewcommand\footnoterule{\vspace*{1pt}%
\color{cream}\hrule width 3.5in height 0.4pt \color{black}\vspace*{5pt}} 
\setcounter{secnumdepth}{5}

\makeatletter 
\renewcommand\@biblabel[1]{#1}            
\renewcommand\@makefntext[1]%
{\noindent\makebox[0pt][r]{\@thefnmark\,}#1}
\makeatother 
\renewcommand{\figurename}{\small{Fig.}~}
\sectionfont{\sffamily\Large}
\subsectionfont{\normalsize}
\subsubsectionfont{\bf}
\setstretch{1.125} 
\setlength{\skip\footins}{0.8cm}
\setlength{\footnotesep}{0.25cm}
\setlength{\jot}{10pt}
\titlespacing*{\section}{0pt}{4pt}{4pt}
\titlespacing*{\subsection}{0pt}{15pt}{1pt}

\fancyfoot{}
\fancyfoot[LO,RE]{\vspace{-7.1pt}\includegraphics[height=9pt]{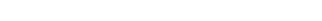}}
\fancyfoot[CO]{\vspace{-7.1pt}\hspace{11.9cm}\includegraphics{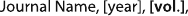}}
\fancyfoot[CE]{\vspace{-7.2pt}\hspace{-13.2cm}\includegraphics{head_foot/RF}}
\fancyfoot[RO]{\footnotesize{\sffamily{1--\pageref{LastPage} ~\textbar  \hspace{2pt}\thepage}}}
\fancyfoot[LE]{\footnotesize{\sffamily{\thepage~\textbar\hspace{4.65cm} 1--\pageref{LastPage}}}}
\fancyhead{}
\renewcommand{\headrulewidth}{0pt} 
\renewcommand{\footrulewidth}{0pt}
\setlength{\arrayrulewidth}{1pt}
\setlength{\columnsep}{6.5mm}
\setlength\bibsep{1pt}

\makeatletter 
\newlength{\figrulesep} 
\setlength{\figrulesep}{0.5\textfloatsep} 

\newcommand{\topfigrule}{\vspace*{-1pt}%
\noindent{\color{cream}\rule[-\figrulesep]{\columnwidth}{1.5pt}} }

\newcommand{\botfigrule}{\vspace*{-2pt}%
\noindent{\color{cream}\rule[\figrulesep]{\columnwidth}{1.5pt}} }

\newcommand{\dblfigrule}{\vspace*{-1pt}%
\noindent{\color{cream}\rule[-\figrulesep]{\textwidth}{1.5pt}} }

\makeatother

\twocolumn[
  \begin{@twocolumnfalse}
{\includegraphics[height=30pt]{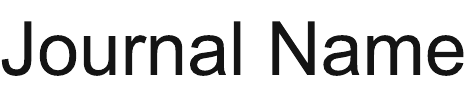}\hfill\raisebox{0pt}[0pt][0pt]{\includegraphics[height=55pt]{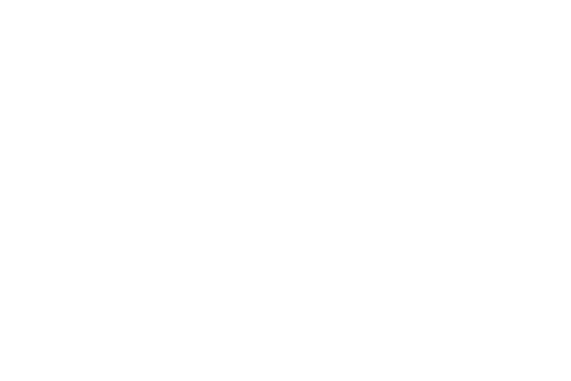}}\\[1ex]
\includegraphics[width=18.5cm]{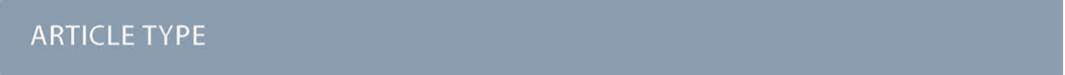}}\par
\vspace{1em}
\sffamily
\begin{tabular}{m{4.5cm} p{13.5cm} }

\includegraphics{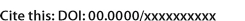} & \noindent\LARGE{\textbf{Magnetism Induced by Azanide and Ammonia Adsorption in Defective Molybdenum Disulfide and Diselenide: A First-Principles Study}}\\
\vspace{0.3cm} & \vspace{0.3cm} \\

 & \noindent\large{Guilherme S. L. Fabris,\textit{$^{a}$} Bruno Ipaves,\textit{$^{a}$} Raphael B. Oliveira,\textit{$^{a,b}$} Humberto R. Gutierrez\textit{$^{c}$}, Marcelo L. Pereira Junior,\textit{$^{b,d}$} and Douglas S. Galvão\textit{$^{a,\ast}$}}\\

\includegraphics{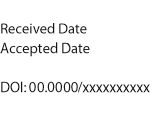} & \noindent\normalsize{Two-dimensional (2D) transition metal dichalcogenides (TMDs) have attracted considerable attention due to their tunable structural, electronic, and spin-related properties, particularly in the presence of point defects and molecular adsorbates. Motivated by these aspects, we have investigated using first-principles methods the magnetic properties induced by azanide (NH$_2$) and ammonia (NH$_3$) adsorption on defective monolayers of Molybdenum Disulfide (MoS$_2$) and Diselenide(MoSe$_2$). Spin-polarized density functional theory (DFT) was employed to investigate the impact of mono- and di-vacancies on the local spin environment and the role of molecular adsorption in modifying magnetic behavior. The results show that pristine chalcogen vacancies do not generate magnetism, whereas the adsorption of NH$_2$ and NH$_3$ creates localized magnetic moments in Mo-based dichalcogenides. A notable case occurs for MoSe$_2$, where NH$_3$ dissociation into NH$_2$ and H fragments on the same side of the surface produces a net magnetic moment of 2.0 $\mu_B$. Tests performed on W-based dichalcogenides under equivalent conditions showed no magnetic response, and are reported here only for comparison. These findings demonstrate that molecular adsorption combined with defect engineering can be a practical approach to tune magnetism in 2D materials, with potential relevance for spintronic and sensing applications.} \\

\end{tabular}

 \end{@twocolumnfalse} \vspace{0.6cm}

  ]

\renewcommand*\rmdefault{bch}\normalfont\upshape
\rmfamily
\section*{}
\vspace{-1cm}

\footnotetext{\textit{$^{a}$~Applied Physics Department and Center for Computational Engineering \& Sciences, State University of Campinas, Campinas, São Paulo 13083-970, Brazil.}}

\footnotetext{\textit{$^{b}$~Department of Materials Science and NanoEngineering, Rice University, Houston, Texas 77005, United States.}}

\footnotetext{\textit{$^{c}$~Department of Physics, University of South Florida 4202 E Fowler Ave, Tampa, Florida 33620, USA}}

\footnotetext{\textit{$^{d}$~Department of Electrical Engineering, College of Technology, University of Brasília, Brasília, Federal District 70910-900, Brazil}}

\footnotetext{\textit{$^{\ast}$~Corresponding Author: galvao@ifi.unicamp.br}}

 
\section{Introduction}

Two-dimensional (2D) transition metal dichalcogenides (TMDs) have attracted considerable attention due to their tunable electronic \cite{electronic1,electronic2}, optical \cite{vu2019electronic,MoS2}, and structural \cite{ipaves2025unraveling,joseph2023review} properties, which make them relevant for diverse technological applications \cite{joseph2023review}. Beyond their intrinsic characteristics, the behavior of TMD monolayers can be significantly modified by the presence of point defects, such as mono and divacancies, as well as by the adsorption of gas molecules \cite{haldar2015systematic,devi2023ferromagnetism,de2024augmented,Lilia1,Lilia2}. Previous studies indicate that vacancies of chalcogen atoms (S or Se) typically do not induce spin density variations in Mo-based dichalcogenides. In contrast, metal vacancies can give rise to localized magnetic moments \cite{yang2019electronic,gaikwad2025long}.

The role of molecular adsorption in defective TMDs has been examined for species such as H$_2$O, O$_2$, and O$_3$, which can stabilize near vacancy sites and, in some cases, dissociate to alter the local electronic structure \cite{ataca2012dissociation,devi2023ferromagnetism}. In parallel, defects such as vacancies and interstitials, which are inevitably formed during material growth, may generate local magnetic moments capable of interacting over long ranges \cite{gaikwad2025long}. These findings underline the importance of modifications in the spin density environment and suggest that the impact of adsorbates depends strongly on the type of vacancy and the surrounding chemical environment.

Although progress has been made, systematic investigations of small adsorbates such as NH$_2$ and NH$_3$ in Mo-based TMDs remain limited. In particular, the combined effects of mono and divacancies, as well as multiple adsorbates, and the resulting distribution of local magnetic moments are still insufficiently explored. Addressing this knowledge gap is important for understanding spin distributions in 2D TMDs and for advancing the design of sensor and spintronic applications.

In this work, we have carried out first-principles density functional theory (DFT) simulations to investigate the effects of azanide (NH$_2$) and ammonia (NH$_3$) adsorption on monolayers of MoS$_2$ and MoSe$_2$ containing mono- and divacancies. We analyzed how adsorbates modify the local spin densities and identified conditions under which magnetism can be induced in otherwise non-magnetic defective chalcogens. For comparison, additional tests on WS$_2$ and WSe$_2$ showed no magnetic response under equivalent conditions. These results provide insights into adsorbate-driven spin distortions in Mo-based dichalcogenides and contribute to a better understanding of defect-engineered 2D materials with tunable spin properties.

\section{Methodology}

To investigate the structural, electronic, and spin densities of MoX$_2$ ($\text X = \text S$ or Se) with mono and divacancies, as well as the effect of NH$_3$ molecules on their properties, we performed \textit{ab initio} simulations based on DFT~\cite{kohn1965self} as implemented in the SIESTA code~\cite{soler2002siesta,garcia2020siesta}. The exchange-correlation effects were described using the PBE (Perdew-Burke-Ernzerhof) functional~\cite{perdew1996generalized}, combined with a DZP (double-$\zeta$ polarization) basis set composed of numerical atomic orbitals. A real-space mesh cutoff of 400~Ry and a $\Gamma$-centered Monkhorst-Pack grid~\cite{monkhorst1976special} of $8\times8\times1$ and $4\times4\times1$ $k$-points were adopted for the monolayer unit cell and the $3\times3\times1$ supercell, respectively. A vacuum region of 25~\r{A} was included along the perpendicular direction to avoid spurious interactions between periodic images. The convergence threshold for the self-consistent field cycle was set to $10^{-4}$ for the density matrix tolerance, and the structural relaxation was performed until the residual forces were smaller than 0.05~eV/\r{A}. All calculations were carried out within a spin-polarized framework.

\section{Results}

Initially, we investigated the changes in the spin densities of MoS$_2$ and WSe$_2$ monolayers in the presence of defects. As a first step, we optimized the pristine MoX$_2$ structures to validate the accuracy of the computational setup. The optimized lattice parameters were $a = b = 3.17$~\r{A} for MoS$_2$ and $a = b = 3.31$~\r{A} for MoSe$_2$, with corresponding Mo-S and Mo-Se bond lengths of 2.42~\r{A} and 2.54~\r{A}. These values are consistent with those reported in the literature, confirming the reliability of the adopted methodology~\cite{haldar2015systematic,shafqat2017dft,zhang2015strain,devi2023ferromagnetism}.

\begin{figure*}[t!]
    \centering
    \includegraphics[width=0.65\linewidth]{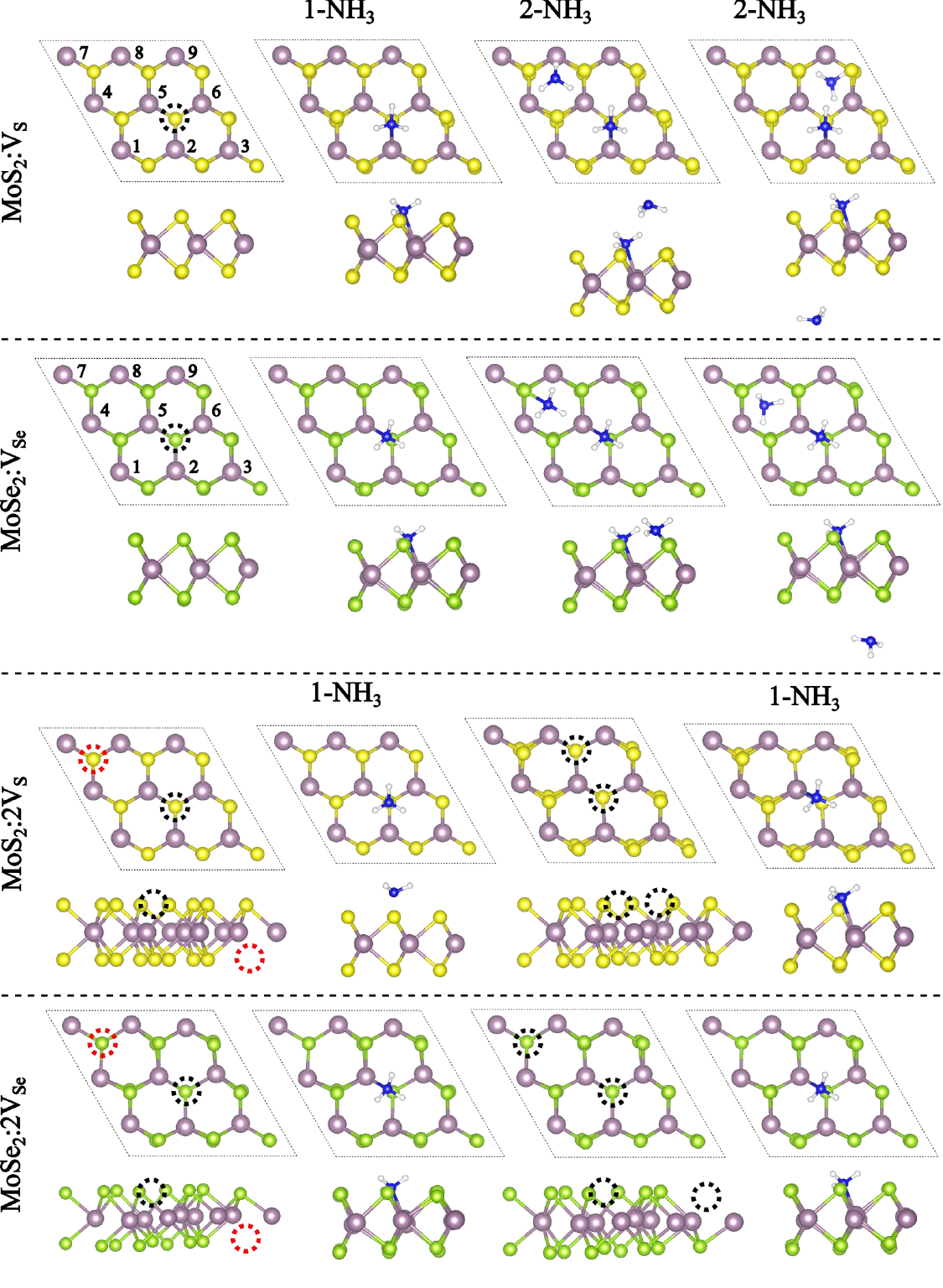}
    \caption{Schematic illustration of the top and side views of a $3\times3\times1$ supercell of MoX$_2$ ($\text X =  \text S$ or Se) featuring one vacancy (VX) and two vacancies (2VX). The figure includes cases investigated with the adsorption of one NH$_3$ molecule and two NH$_3$ molecules. Dashed black and red circles indicate the vacancy positions, while the numbers 1--9 correspond to the label of the Mo atoms referenced in all Tables.}
    \label{fig1}
\end{figure*}

After validating the computational parameters, we created a $3\times3\times1$ supercell of MoX$_2$ (X = S or Se) and introduced mono- and divacancy models. The divacancy was considered in two configurations: both vacancies on the same side of the monolayer and one vacancy on each side of the supercell (alternated), as illustrated in Figure~\ref{fig1}. These defective models were then used to investigate the influence of NH$_3$ adsorption at the vacancy sites, with representative configurations shown in Figure~\ref{fig1}.

The creation of S and Se vacancies resulted in defective structures without spin density variations, in agreement with previous reports that describe the absence of magnetism in such systems~\cite{haldar2015systematic,shafqat2017dft,zhang2015strain,devi2023ferromagnetism}. The adsorption of H$_2$O and NH$_3$ in Mo-based TMDs has been examined in the literature, and most studies indicate negligible changes in the spin environment~\cite{devi2023ferromagnetism,li2016markedly,de2024augmented}. Nevertheless, the dissociation of H$_2$O at vacancy sites has been reported to induce significant modifications in the local spin distribution~\cite{devi2023ferromagnetism,ataca2012dissociation}.

For the NH$_3$ adsorption, a single molecule stabilizes near the vacancy site at distances of 2.38~\r{A} and 2.36~\r{A} from the nearest Mo atom in MoS$_2$ and MoSe$_2$, respectively. When two NH$_3$ molecules are present, one occupies a similar position to the single-molecule case. At the same time, the second stabilizes farther from the surface, with distances from the nearest S or Se atoms ranging between 1.79~\r{A} and 2.94~\r{A}.

\begin{figure}[htb!]
    \centering
    \includegraphics[width=0.95\linewidth]{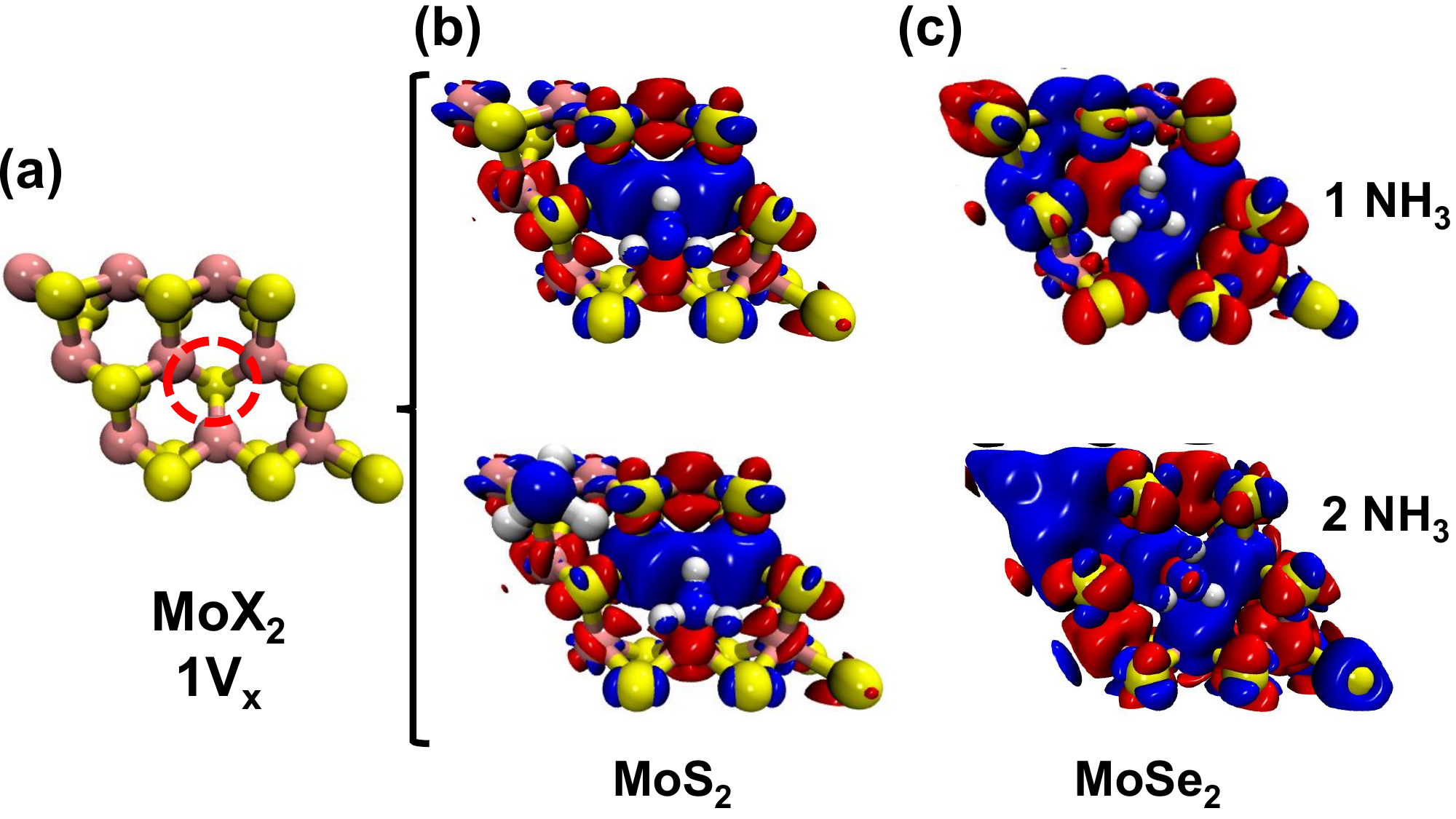}
    \caption{Spin density maps for (a) MoX$_2$ monolayers with a monovacancy, (b) MoS$_2$, and (c) MoSe$_2$ with one and two adsorbed NH$_3$ molecules. The dashed red circle indicates the position of the vacancy. In the maps, red color indicates spin-up density, while blue represents spin-down density.}
    \label{fig2}
\end{figure}

\begin{figure}[htb!]
    \centering
    \includegraphics[width=0.95\linewidth]{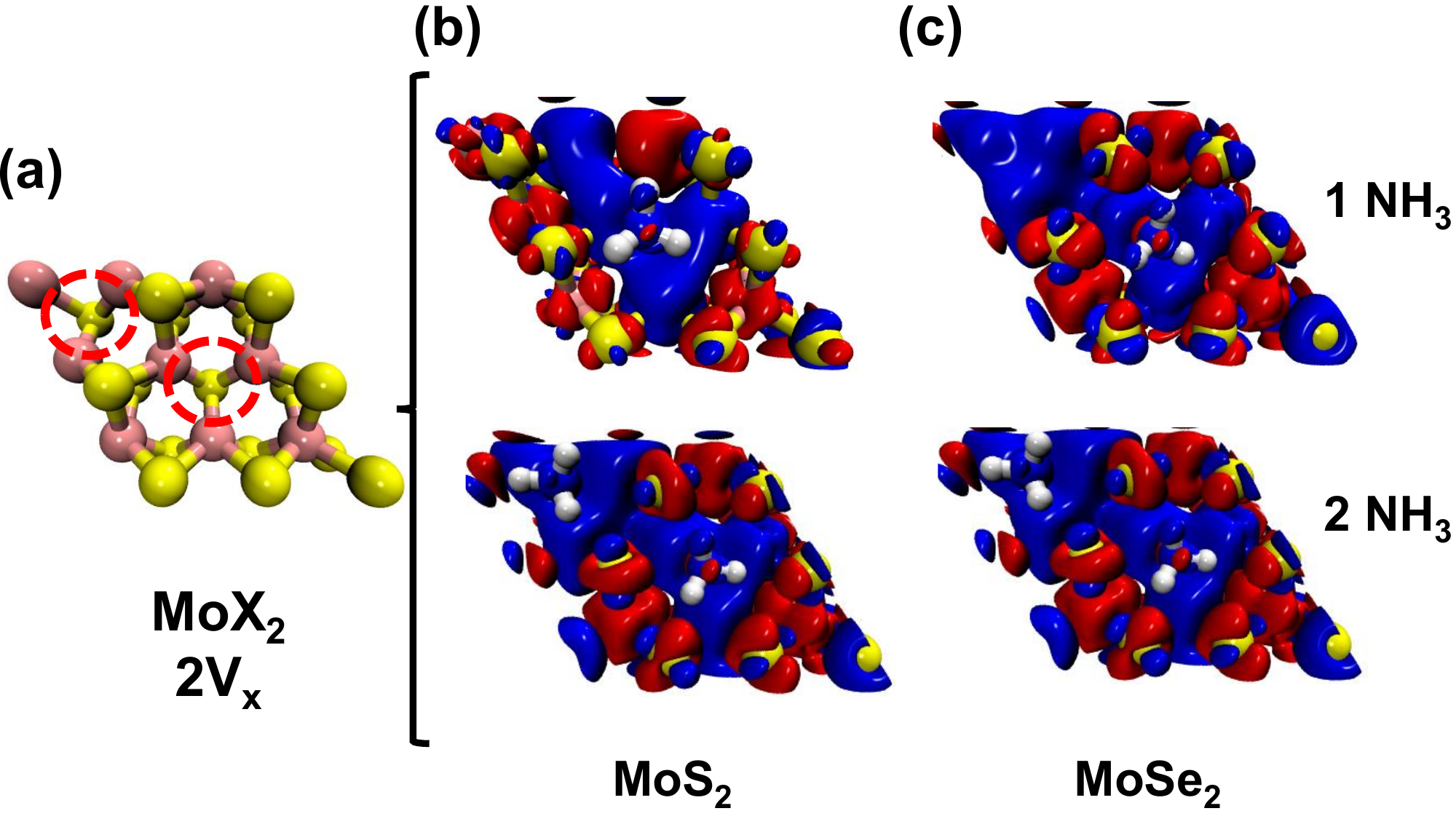}
    \caption{Spin density maps for (a) MoX$_2$ monolayers with a divacancy, (b) MoS$_2$, and (c) MoSe$_2$ with one and two adsorbed NH$_3$ molecules. The dashed red circle indicates the position of the vacancy. In the maps, red color indicates spin-up density, while blue represents spin-down density.}
    \label{fig3}
\end{figure}

A key result of this study is that NH$_3$ adsorption increases the local magnetic moment of Mo atoms near vacancy sites in both MoS$_2$ and MoSe$_2$ (Figures~\ref{fig2} and~\ref{fig3}, Tables~\ref{tab1} and~\ref{tab2}). This effect, rarely reported in the literature, was observed at specific Mo atoms surrounding the adsorption site. In the case of a divacancy on the same side of the monolayer, the increase became more pronounced, with additional Mo atoms developing non-zero magnetic moments. Quantitatively, MoS$_2$ exhibited an enhancement of approximately 21.3\%, while MoSe$_2$ showed a much larger increase of 200\%. By contrast, the alternated divacancy configuration did not modify the response, yielding values comparable to those of the single-vacancy case. These results indicate that exposure to NH$_3$ can enhance induced magnetism in Mo-based dichalcogenides, as illustrated by the spin density difference maps in Figures~\ref{fig2} and~\ref{fig3}.

\begin{table*}[htb!]
    \centering
    \caption{Spin-polarized electron distribution for Mo atoms near vacancy sites with one NH$_3$ adsorption in MoS$_2$ and MoSe$_2$. The Table presents the local magnetic moments ($\mu_\text{Local}$) in units of $e^-$ at each Mo atom (labeled according to Figure~\ref{fig1}), calculated as the difference in spin-up and spin-down electron populations. Notably, Mo atoms at specific positions exhibit non-zero resulting magnetic moments, with $\mu_\text{Local}$ values possibly indicating magnetic response. This effect intensifies with divacancy configurations, as indicated by the increased magnetic moments on other Mo atoms.}
    \label{tab1}
    \vspace{0.5em}
    \begin{tabular}{c|cc|cc}
        \hline
        \textbf{Mo Atom Nº} & \multicolumn{2}{c|}{\textbf{MoS$_2$}} & \multicolumn{2}{c}{\textbf{MoSe$_2$}} \\
        & $\mu_\text{Local}$ (1 V$_\text{S}$) & $\mu_\text{Local}$ (2 V$_\text{S}$) & $\mu_\text{Local}$ (1 V$_\text{Se}$) & $\mu_\text{Local}$ (2 V$_\text{Se}$) \\
        \hline
        1 & -0.058 & -0.133 &  0.065 & -0.984 \\
        2 & -0.222 &  1.158 &  2.098 &  1.265 \\
        3 & -0.058 & -0.232 & -1.079 & -0.758 \\
        4 & -0.104 & -0.265 &  0.472 &  2.196 \\
        5 &  1.418 &  1.287 & -1.474 &  1.410 \\
        6 &  1.417 &  1.049 &  2.098 &  1.265 \\
        7 &  0.026 & -0.075 & -0.433 &  1.303 \\
        8 &  0.025 &  1.067 &  0.472 &  2.196 \\
        9 & -0.332 & -1.150 &  0.065 & -0.984 \\
        \hline
    \end{tabular}
\end{table*}

\begin{table*}[htb!]
    \centering
    \caption{Spin-polarized electron distribution for Mo atoms near vacancy sites with two NH$_3$ adsorptions in MoS$_2$ and MoSe$_2$ in different positions. The Table presents the Mo local magnetic moments ($\mu_\text{Local}$) in $e^-$, illustrating the resulting magnetic moment response (atom numbers labeled according to Figure~\ref{fig1}). When the molecules were positioned on opposite sides for MoSe$_2$, the $\mu_\text{Local}$ values resemble those from the single vacancy and one NH$_3$ case (see Table~\ref{tab1}).}
    \label{tab2}
    \vspace{0.5em}
    \begin{tabular}{c|cc|cc}
        \hline
        \textbf{Mo Atom Nº} & \multicolumn{2}{c|}{\textbf{MoS$_2$}} & \multicolumn{2}{c}{\textbf{MoSe$_2$}} \\
        & $\mu_\text{Local}$ (Same side) & $\mu_\text{Local}$ (Opposite side) & $\mu_\text{Local}$ (Same side) & $\mu_\text{Local}$ (Opposite side) \\
        \hline
        1 & -0.058 & -0.058 & -0.369 &  0.070 \\
        2 & -0.226 & -0.227 &  1.933 &  2.101 \\
        3 & -0.060 & -0.058 & -0.612 & -1.089 \\
        4 & -0.102 & -0.104 & -0.697 &  0.447 \\
        5 &  1.424 &  1.415 &  1.655 & -1.481 \\
        6 &  1.422 &  1.410 &  1.925 &  2.104 \\
        7 &  0.024 &  0.031 & -0.681 & -0.409 \\
        8 &  0.027 &  0.032 & -0.440 &  0.463 \\
        9 & -0.340 & -0.330 & -0.503 &  0.049 \\
        \hline
    \end{tabular}
\end{table*}

We further investigated the resulting magnetic moment behavior induced by the adsorption of two NH$_3$ molecules on MoS$_2$ and MoSe$_2$ surfaces with a single vacancy. For both MoS$_2$ and MoSe$_2$, placing two NH$_3$ molecules on the same side of the surface resulted in a non-zero magnetic moment, which was also observed when positioning one NH$_3$ molecule on each side of the surface. In the case of MoS$_2$, the magnetic moment behavior was similar to that observed in configurations with a single vacancy and one NH$_3$ molecule. Similarly, for MoSe$_2$, placing one NH$_3$ on each side of the surface produced a non-magnetic moment response comparable to that observed in configurations with a single vacancy and one NH$_3$ (see Tables~\ref{tab1} and~\ref{tab2}).

\begin{figure}[htb!]
    \centering
    \includegraphics[width=0.95\linewidth]{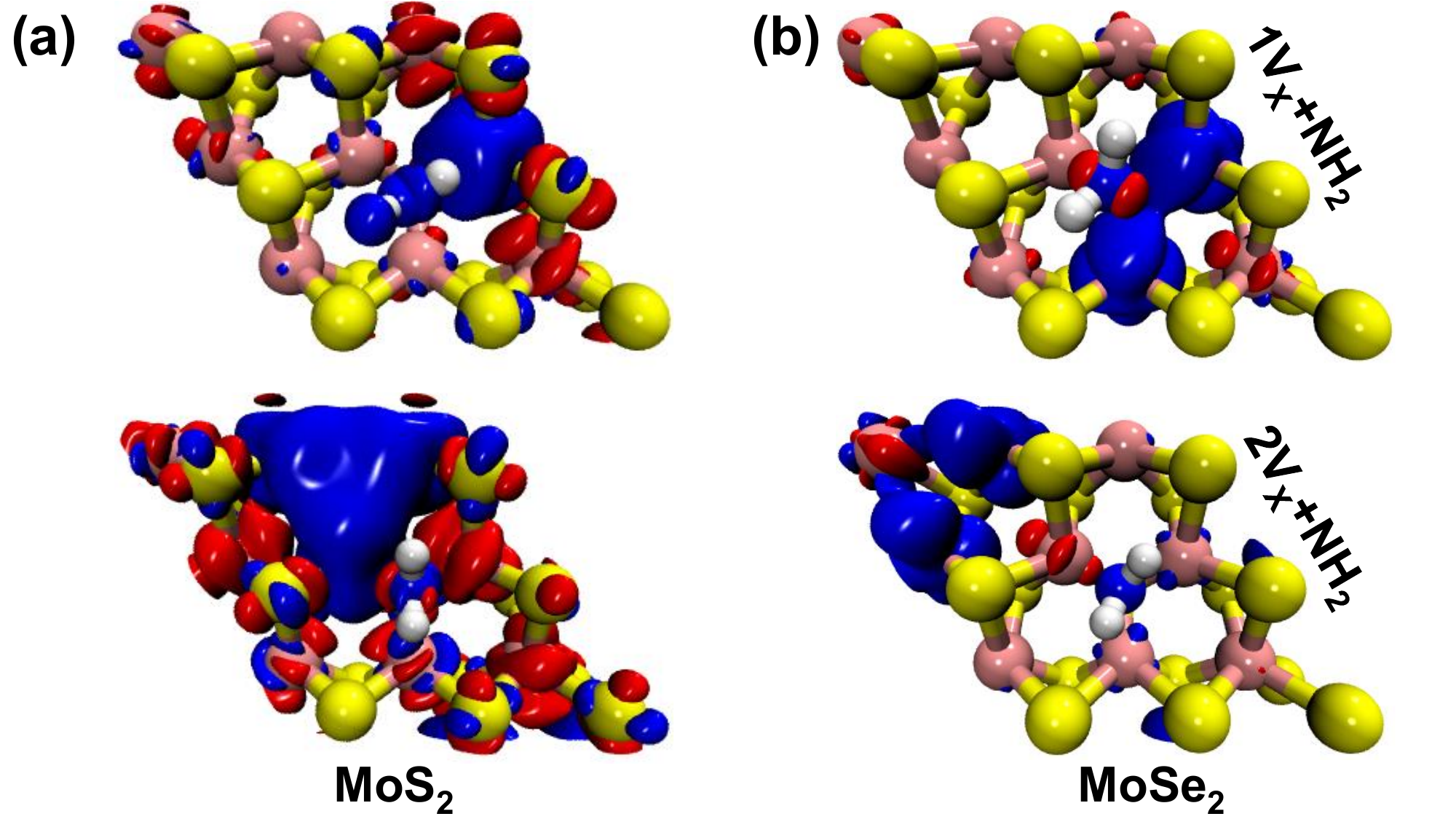}
    \caption{Spin density maps for (a) MoS$_2$ and (b) MoSe$_2$ monolayers with one and two vacancies (top and bottom figures, respectively), and with one adsorbed NH$_2$ molecule.}
    \label{fig4}
\end{figure}

We have extended our investigation to the adsorption of NH$_2$ molecules on MoS$_2$ and MoSe$_2$ surfaces with mono and di-vacancies. The results, presented in Figure~\ref{fig4}, Tables~\ref{tab3} and~\ref{tab4}, reveal distinct magnetic moment behaviors compared to the NH$_3$ case, providing further insights into the role of adsorbates and vacancy configurations in spin density variations.

For NH$_2$ adsorption, we observed a magnetic response in Mo atoms near the vacancy sites for both MoS$_2$ and MoSe$_2$. From the simulations, it is evident that the magnetic moment in MoS$_2$ increases substantially when a divacancy is introduced; however, this behavior was not observed in MoSe$_2$. In MoS$_2$, significant magnetic moments appeared at specific Mo atoms, with local magnetic moments ($\mu_\text{Local}$) reaching up to 0.911~$e^-$ for the single vacancy, and showing only slight variations under divacancy conditions. Conversely, in MoSe$_2$, the magnetic moments were generally weaker, with values below 0.5~$e^-$ in most cases (Table~\ref{tab3}).

These findings suggest that NH$_2$ adsorption alone is sufficient to induce an increase in the resulting magnetic moment of Mo atoms, but that the effect does not scale with increasing vacancy density, as observed in the NH$_3$ case. This suggests that NH$_2$ molecules interact differently with the MoX$_2$ surface, possibly due to their smaller size and distinct electronic configuration compared to NH$_3$, resulting in distinct bonding characteristics and charge redistribution near the vacancy sites. The absence of a significant enhancement in the total magnetic moment for divacancy configurations further emphasizes the localized nature of the magnetic response induced by NH$_2$.

\begin{table*}[htb!]
    \centering
    \caption{Spin-polarized electron distribution for Mo atoms near vacancy sites with one NH$_2$ adsorption in MoS$_2$ and MoSe$_2$. The Table presents the local magnetic moments ($\mu_\text{Local}$) in $e^-$ at each Mo atom (labeled according to Figure~\ref{fig1}), calculated by the difference in spin-up and spin-down electron populations. Notably, Mo atoms at specific positions exhibit induced magnetic moments, with $\mu_\text{Local}$ values highlighting the magnetic moment response. Unlike the NH$_3$ case, this effect does not increase with divacancy configurations.}
    \label{tab3}
    \vspace{0.5em}
    \begin{tabular}{c|cc|cc}
        \hline
        \textbf{Mo Atom Nº} & \multicolumn{2}{c|}{\textbf{MoS$_2$}} & \multicolumn{2}{c}{\textbf{MoSe$_2$}} \\
        & $\mu_\text{Local}$ (Vacancy) & $\mu_\text{Local}$ (Divacancy) & $\mu_\text{Local}$ (Vacancy) & $\mu_\text{Local}$ (Divacancy) \\
        \hline
        1 &  0.056 & -0.032 &  0.000 &  0.002 \\
        2 &  0.032 &  0.692 &  0.041 &  0.020 \\
        3 & -0.045 & -0.051 & -0.028 & -0.008 \\
        4 & -0.026 &  0.008 &  0.007 &  0.482 \\
        5 &  0.032 &  0.012 &  0.036 & -0.050 \\
        6 &  0.911 &  0.056 &  0.934 &  0.023 \\
        7 & -0.026 & -0.010 &  0.009 & -0.033 \\
        8 &  0.049 &  0.095 &  0.003 &  0.508 \\
        9 & -0.044 &  0.188 & -0.026 &  0.002 \\
        \hline
    \end{tabular}
\end{table*}

\begin{figure}[htb!]
    \centering
    \includegraphics[width=1\linewidth]{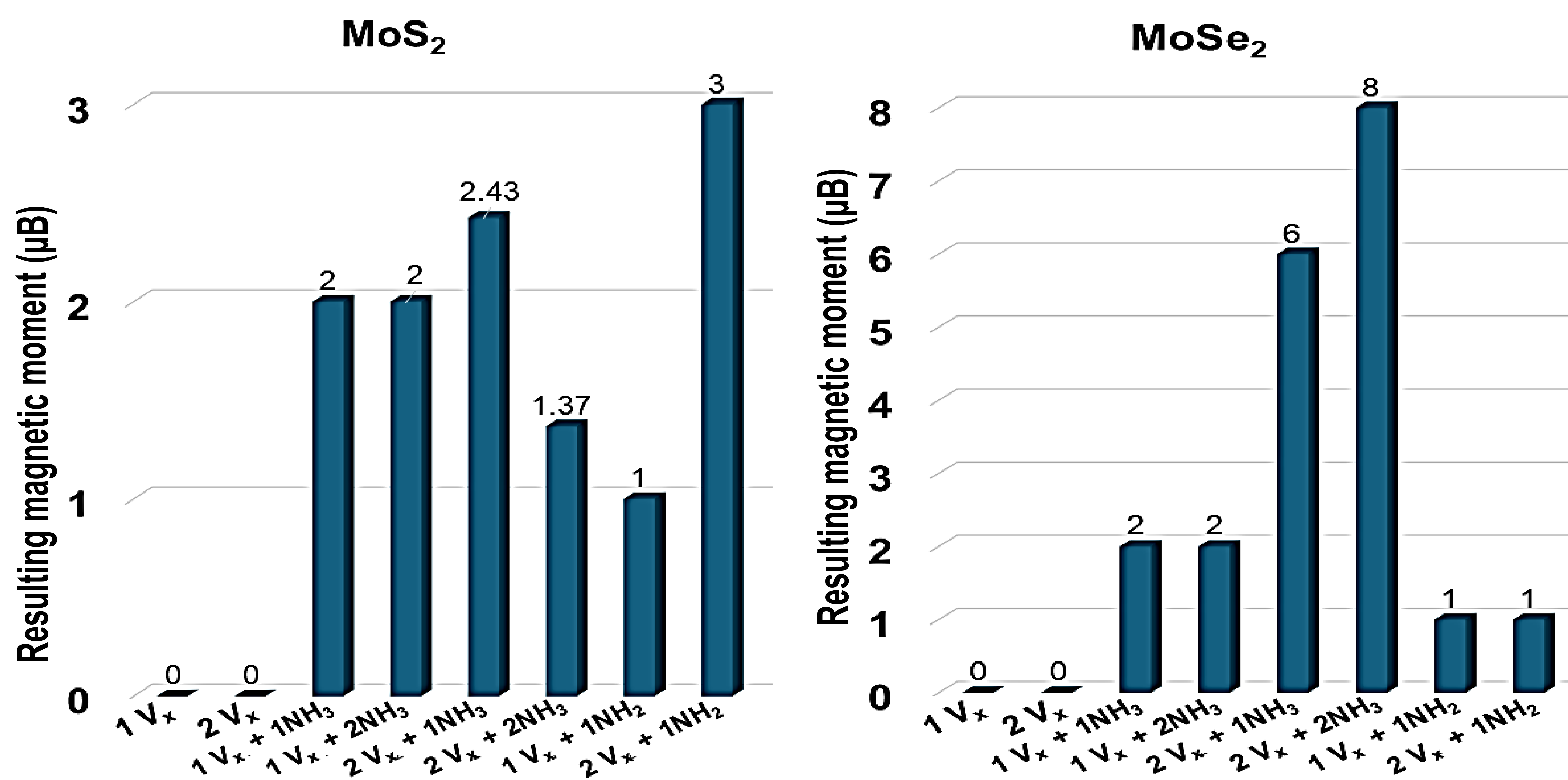}
    \caption{The resulting magnetic moment of each case considered in this work. The results reflect the values obtained from the same side vacancy and NH$_3$/NH$_2$ molecule adsorption.}
    \label{fig5}
\end{figure}

Following this, we further investigated the magnetic moment behavior induced by two NH$_2$ molecules adsorbed on MoS$_2$ and MoSe$_2$ surfaces with a single vacancy. For MoS$_2$, adsorbing two NH$_2$ molecules on the surface resulted in local magnetic moments distributed among Mo atoms near the vacancy site (Table~\ref{tab4}). For MoSe$_2$, positioning one NH$_2$ molecule on each side of the surface induced a magnetic moment; however, placing two NH$_2$ molecules on the same side of the surface resulted in a zero magnetic moment. To better visualize the resulting magnetic moment of most of our results, we compiled the values in Figure~\ref{fig5}.

We also investigated the possible dissociation of NH$_3$ into NH$_2$ and H fragments adsorbed on the surface, considering two distinct configurations: both species located on the same side of the exposed surface, and an alternating arrangement with NH$_2$ and H positioned on opposite sides (top and bottom), which the results can be seen in Table \ref{tab5}. For MoS$_2$, a net magnetic moment of 0.444 $\mu_B$ was observed in both configurations, indicating that the dissociation induces magnetization regardless of the spatial distribution of the fragments. In contrast, for MoSe$_2$, a net magnetic moment of 2.0 $\mu_B$ was only observed when both NH$_2$ and H were adsorbed on the same side of the surface. At the same time, the alternating configuration did not result in any net magnetization.

These results suggest that the magnetic response of the system is highly sensitive to both the nature of the chalcogen atom and the spatial arrangement of the dissociated species. The stronger magnetization observed in MoSe$_2$ under specific adsorption geometry may be attributed to enhanced spin polarization effects mediated by the heavier selenium atoms and the localized electronic interactions between co-adsorbed fragments. This highlights the potential for tuning magnetic properties in 2D materials via controlled molecular dissociation and adsorption configurations.

\begin{table*}[htb!]
    \centering
    \caption{Spin-polarized electron distribution for Mo atoms near vacancy sites with two NH$_2$ adsorptions in MoS$_2$ and MoSe$_2$. The Table presents the Mo local magnetic moments ($\mu_\text{Local}$) in $e^-$, illustrating the magnetic response (atom numbers labeled according to Figure~\ref{fig1}). We found no magnetism in MoSe$_2$ for the configuration with both NH$_2$ molecules on the same side.}
    \label{tab4}
    \vspace{0.5em}
    \begin{tabular}{c|cc|cc}
        \hline
        \textbf{Mo Atom Nº} & \multicolumn{2}{c|}{\textbf{MoS$_2$}} & \multicolumn{2}{c}{\textbf{MoSe$_2$}} \\
        & $\mu_\text{Local}$ (Same side) & $\mu_\text{Local}$ (Opposite side) & $\mu_\text{Local}$ (Same side) & $\mu_\text{Local}$ (Opposite side) \\
        \hline
        1 &  0.048 &  0.022 & 0.000 & 0.197 \\
        2 &  0.029 &  0.040 & 0.000 & 0.029 \\
        3 & -0.045 & -0.022 & 0.000 & -0.036 \\
        4 & -0.029 &  0.006 & 0.000 & -0.008 \\
        5 &  0.030 &  0.043 & 0.000 & 0.008 \\
        6 &  0.910 &  0.825 & 0.000 & 0.973 \\
        7 & -0.022 & -0.007 & 0.000 & 0.032 \\
        8 &  0.040 &  0.011 & 0.000 & -0.002 \\
        9 & -0.043 & -0.024 & 0.000 & 0.007 \\
        \hline
    \end{tabular}
\end{table*}

\begin{table*}[htb!]
    \centering
    \caption{Spin-polarized electron distribution for Mo atoms near vacancy sites with one NH$_2$ and one H adsorption in MoS$_2$ and MoSe$_2$. The Table presents the local magnetic moments ($\mu_\text{Local}$) in $e^-$ at each Mo atom (labeled according to Figure~\ref{fig1}), calculated by the difference in spin-up and spin-down electron populations. Notably, Mo atoms at specific positions exhibit induced magnetic moments, with $\mu_\text{Local}$ values highlighting the magnetic moment response. }
    \label{tab5}
    \vspace{0.5em}
    \begin{tabular}{c|cc|cc}
        \hline
        \textbf{Mo Atom Nº} & \multicolumn{2}{c|}{\textbf{MoS$_2$}} & \multicolumn{2}{c}{\textbf{MoSe$_2$}} \\
        & $\mu_\text{Local}$ (Same side) & $\mu_\text{Local}$ (Opposite side) & $\mu_\text{Local}$ (Same side) & $\mu_\text{Local}$ (Opposite side) \\
        \hline
        1 & -0.005 &  0.000 &  0.064 &  0.000 \\
        2 & -0.011 & -0.018 &  0.042 &  0.000 \\
        3 & -0.026 & -0.034 &  0.003 &  0.000 \\
        4 & -0.038 & -0.047 &  0.678 &  0.000 \\
        5 & -0.014 & -0.019 &  0.023 &  0.000 \\
        6 &  0.601 &  0.620 &  0.922 &  0.000 \\
        7 & -0.038 & -0.047 &  0.015 &  0.000 \\
        8 & -0.041 & -0.021 & -0.001 &  0.000 \\
        9 & -0.025 & -0.035 & -0.029 &  0.000 \\
        \hline
    \end{tabular}
\end{table*}

We extended our investigation to W-based dichalcogenides, WS$_2$ and WSe$_2$, using the same approach used to MoS$_2$ and MoSe$_2$. Initially, we explored the magnetic moment behavior of the pristine monolayers as well as systems with mono and di-vacancies (V$_\text{X}$ and 2V$_\text{X}$, respectively). Unlike MoX$_2$, no resulting magnetic moment was observed in WS$_2$ or WSe$_2$ for any vacancy configuration, with the magnetic moments consistently obtained as zero.

Additionally, we examined the adsorption of one and two NH$_3$ molecules on the vacancy sites of WS$_2$ and WSe$_2$. In all cases, the systems remained with a zero resulting magnetic moment, indicating that NH$_3$ adsorption does not induce spin density changes in these materials, even in the presence of vacancies.

Non-zero spin density emerged only when a W atom was removed instead of an S or Se atom. This suggests that the absence of a resulting magnetic moment in the previous configurations arises from the electronic structure of the W atoms and their interaction with the surrounding lattice. This behavior is consistent with previous theoretical studies, which reported that magnetism appears only in the presence of W$_2$ or WSe$_6$ vacancies~\cite{yang2019electronic}.

Our results show that creating S or Se vacancies in MoX$_2$ does not inherently lead to a non-zero resulting magnetic moment; however, the presence of NH$_2$ or NH$_3$ molecules can modify the local environment and lead to a non-zero magnetic moment. This effect becomes more pronounced under conditions of high defect density and low NH$_3$ concentration. On the other hand, no resulting magnetic moment was observed in WX$_2$ systems with either mono- or di-vacancies of X. These findings highlight a significant contrast between the spin density properties of Mo- and W-based dichalcogenides under similar conditions, emphasizing the critical role of the transition metal in determining the resulting magnetic moment behavior of these materials.

\section{Conclusions}

In summary, spin-polarized DFT simulations were employed to investigate the effects of NH$_2$ and NH$_3$ adsorption on defective MoX$_2$ ($\text X = \text S$, Se) monolayers. The results confirm that pristine chalcogen vacancies do not induce magnetism, while molecular adsorption can create localized magnetic moments in Mo-based dichalcogenides. A notable case was observed for MoSe$_2$, where NH$_3$ dissociation into NH$_2$ and H fragments on the same side of the surface produced a net magnetic moment of 2.0 $\mu_B$. For comparison, W-based TMDs were also examined and remained non-magnetic under equivalent conditions. These findings suggest that molecular adsorption, combined with defect engineering, influences the magnetic behavior of Mo-based TMDs, providing insights for future studies on spin-related phenomena in low-dimensional systems.

\section*{Author Contributions}
G.S.L.F.: Methodology, Data Curation, Software, Formal analysis, Investigation, Writing - Original Draft, Writing - Review \& Editing, Visualization. 
B.I.: Methodology, Formal analysis, Investigation, Writing - Original Draft, Writing - Review \& Editing, Visualization
R.B.O.: Methodology, Data Curation, Software, Formal analysis, Investigation, Writing - Original Draft, Writing - Review \& Editing, Visualization.
H.R.G: Conceptualization, Writing - Review \& Editing.
M.L.P.J.: Writing - Original Draft, Funding acquisition, Investigation, Formal Analysis, Writing - Review \& Editing, Visualization.
D.S.G.: Conceptualization, Investigation, Formal Analysis, Resources, Writing - Review \& Editing, Supervision, Project administration.

\section*{Conflicts of interest}

There are no conflicts to declare.

\section*{Acknowledgements}

R.B.O. thanks the National Council for Scientific and Technological Development (CNPq) process numbers 151043/2024-8 and 200257/2025-0. B. I. acknowledges support from CNPq and São Paulo Research Foundation (FAPESP) process numbers 153733/2024-1 and 2024/11016-0. G. S. L. F. acknowledges support from FAPESP process number \#2024/03413-9. M.L.P.J. acknowledges financial support from FAPDF (grant 00193-00001807/2023-16), CNPq (grants 444921/2024-9 and 308222/2025-3), and CAPES (grant 88887.005164/2024-00). D. S. G. acknowledges the Center for Computing in Engineering and Sciences at Unicamp for financial support through the FAPESP/CEPID Grant \#2013/08293-7. We thank the Coaraci Supercomputer Center for computer time (Fapesp grant \#2019/17874-0).
\balance

\bibliography{rsc}
\bibliographystyle{rsc}

\end{document}